\documentclass[12pt,aps,prb,preprint,draft]{revtex4}

\usepackage{longtable}
\setlongtables
\begin{document}

\title{Impact of peer interaction on conceptual test performance}
\author{Chandralekha Singh}
\affiliation{Department of Physics and Astronomy, University of
Pittsburgh, Pittsburgh, Pennsylvania 15260}
%\date{ }

\begin{abstract}
We analyze the effectiveness of working in pairs on the Conceptual
Survey of Electricity and Magnetism test in a calculus-based
introductory physics course. Students who collaborated with a peer
showed significantly larger normalized gain on individual testing than
those who did not collaborate. We did not find statistically
significant differences between the performance of students who were
given an opportunity to formulate their own response before the peer
discussions, compared to those who were not. Peer collaboration also
shows evidence for co-construction of knowledge. Discussions with
individual students show that students themselves value peer
interaction. We discuss the effect of pairing students with different
individual achievements.
\end{abstract}

\maketitle

\section{Introduction}

Cognitive research suggests that an individual must process new
material actively and build proper associations with their prior
knowledge for learning to be meaningful. One way to immerse students
actively in the learning process is to have them interact with each
other. Peer collaboration as a learning tool has been exploited in
many instructional settings and with different types and levels of
student populations.\cite{collaborate} Although the details vary,
students can learn from each other in many different environments.\cite{eric,heller,sokoloff,laws,vygotsky,rogoff,damon,barron,hogan,azmitia} 

In college physics instruction, Eric Mazur\cite{eric} has popularized a
peer instruction method in which the instructor poses several
conceptual multiple-choice questions during the lecture. Students
discuss their reasoning with peers and are polled about their choices.
Mazur has cited several advantages for why this method is effective.
Peer interaction keeps students alert during the lectures because they
know they must discuss the questions with peers, and it also helps them
organize and extend
their knowledge. Articulating one's opinion requires attention to logic
and organization of thought processes. Instant feedback from students
also provides a ``reality check'' to the instructors about the extent
to which students have learned to the concepts. This
check can help instructors adjust the pace of the class
appropriately. Moreover, there often is a mismatch between instructor
and students' expectations of the level of understanding. Peer
instruction helps convey instructor's expectations to the students so
that students can adjust their expectations. Physics
educators also have exploited peer collaboration to teach problem
solving using complex context-rich real-life problems,\cite{heller} to
make lecture demonstrations meaningful to students,\cite{sokoloff} and
to teach physics without lectures in a workshop style.\cite{laws} An
additional advantage of peer collaboration is that it is embedded in a
context that can help students retain and recreate the content by
remembering the discussion.\cite{collaborate}

There have been many more investigations involving the effectiveness of
collaboration in K-12 education, both in science and in other
disciplines. However, there is little quantitative data from
pre-/post-testing (testing before and after group intervention). Most
researchers have analyzed the success of peer collaboration based upon
patterns of student discussion. According to Vygotsky's socio-cultural
perspective,\cite{vygotsky} learning is fundamentally a social process.
In this perspective, student learning can be enhanced by
participation in various social tasks that are designed by 
instructors who are familiar with students' prior knowledge. Many
researchers who are influenced by this perspective investigate the role
of different types of social environment in mediating and facilitating
learning. Rogoff\cite{rogoff} reviewed the research on
collaborative learning in K-12 education. She addressed issues about
the success of peer collaboration based on the analysis of 
individuals and the group as a whole. The analysis of her own research described in her review is not
based upon performance on tests given after collaboration, but on the
interaction between individuals. She found that children are more likely to examine the
logic of arguments with peers than with adults, with more self-generated
clarifications of logic and more commentary during discussions with
peers. Damon\cite{damon} has contended that peer education has an
untapped potential and has advocated peer-based approaches to education.
He has claimed that more expert partners may facilitate skill and
information learning, while peer partners may facilitate conceptual
change or acquisition of new principles. He has asserted that peer
collaboration complements rather than supplants adult teaching, freeing
up teachers' energy and attention and enabling them to focus on
children's other needs. Barron\cite{barron} posed complex multi-step
mathematical problem solving to students and discussed how achieving
coordination, for example, balanced involvement of both individuals in problem solving, are critical for effective collaboration. 

Hogan et al.\cite{hogan} analyzed the discourse patterns and
collaborative scientific reasoning in peer and teacher-guided
discussions and discussed their relative advantages. The
patterns of verbal interaction within peer and teacher-student
scientific sense-making discussions were dissected and the
relation between discourse patterns and sophistication of scientific
reasoning during discussions was studied. It was found that
peers working without a teacher talked more, exhibited a greater
level of reasoning complexity, and were better able to justify arguments
and synthesize information. The presence of a teacher brought students
to a resolution of ideas more efficiently, resulting in a reduced need
for talk and reasoning complexity. No data was provided about the
learning outcomes or the quality of the students' final models. Azmitia
et al.\cite{azmitia} found that when friends were grouped together they were
more likely to spontaneously justify their solutions, check their
answers, and engage in conflicts that were resolved by discussions than groups with members who did not know each other from before.

\section{Background}
In this paper we report on the effectiveness of working in pairs in a
calculus-based introductory physics courses. We chose the conceptual
multiple-choice test, the Conceptual Survey of Electricity and
Magnetism (CSEM),\cite{csem} because it is a standardized test that
covers a wide variety of qualitative (conceptual) problems that are
covered in calculus-based introductory physics courses. Each of the 32
questions on this standardized test has 5 choices (a correct answer and
four distracters). Thus, a score significantly better than 20\% shows a
non-random response. Our investigation supplements previous studies on
peer collaboration discussed in the previous section.
Students had instruction in the relevant physics principles and
concepts before taking the test. Our interest is in exploring the learning gain due to group
interactions during qualitative problem solving. In particular, we want
to know the extent to which group performance differs from individual
performance. We also are interested in understanding the extent to which
students can co-construct knowledge,\cite{collaborate} that is, are
there instances in which the group members collectively choose the
correct response even though each of them had chosen an incorrect
response earlier? Also of interest is the extent to which peers
retained what they have learned. Are there any differences in learning
gains and retention if students are first given an opportunity to think
about the problems individually or if a student with a high initial
individual score before group intervention is paired with another
student with a high or low individual score?

The investigation was performed in four introductory physics courses. 
Two of these courses formed the experimental group in which group
intervention was included, and the other two courses formed the control
group with no group intervention.  The experimental group
consisted of two  types of interventions  to explore the extent to
which giving students an opportunity to formulate their own responses
before  group intervention (IG protocol)  enhanced learning compared
to the case where students first worked in groups without a prior
opportunity to formulate their own responses (GI protocol).  Regardless of the protocol, all
students took the CSEM test individually two weeks later after the initial administration of the test
to assess the extent to which they had retained the relevant
concepts.  Each test counted for one quiz.

Some studies show that heterogenous groups are more effective for group
learning, while others show that working with friends has special
advantages.\cite{azmitia} In our study, students were allowed to
choose their own partners. They were encouraged to discuss the
questions with each other. 
Apart from the explicit encouragement by the instructor, students had an additional motivation to discuss the concepts, because the forthcoming examination covered
the same material. Moreover, students already had extensive experience
working in groups of three in the recitation on context-rich
problems\cite{heller} and in pairs during lecture on Mazur-style
concept tests.\cite{eric} We note that the peer collaboration was
unguided in that there was no help or facilitation from the instructor
except that the physics principles and concepts were covered in earlier
classes.

\section{Discussion of different protocols}

Students in both experimental groups were given the CSEM test twice
during a double class period (110 minutes). To obtain two random
equivalent samples, all students in the ``experimental group'' class
sitting on one side of the aisle worked individually,  followed by
group work, while those on the other side of the aisle worked in groups
first before working individually. In the IG  experimental group
, students first worked individually, and then in groups of two. In the
GI group (group followed by individual), students first worked in
groups of two and then individually. Students worked individually or in
a group for 50 minutes. Between the individual and group testing there
was a short break, and students were required to turn in their first
response so that they could not refer to it when working in pairs (IG)
or individually (GI) the second time. The test answers were not 
discussed with students, so when they switched from individual to group
(or vice versa), they did not know if their initial responses were
correct.

Each class in the IG protocol had 74 students or 37 pairs (for a total
of 148 students). In the GI protocol, one class had 54 students or 27
pairs and another class had 30 students or 15 pairs (total 84
students). As noted, the main motivation for giving the test in both
ways was to assess the effect of thinking individually before peer
discussion. In Mazur-style peer instruction, students are first asked
to think about the concepts individually before talking to peers.
Another reason for using both the IG and GI protocols was to evaluate any
``test-retest'' (also known as ``practice'' or ``carry-over'') effect.

Although the trends in some individual test questions are interesting,
we mostly focus here on the effect of group work on overall test scores.
Table~1 shows the average individual and group scores for the GI and IG
protocols. The average performance on the IG and GI protocols are
statistically indistinguishable, which suggests that giving students an
opportunity to think alone before peer discussion did not improve their
group performance significantly. In the IG protocol, the group
performance after the individual performance is not a ``test-retest''
effect. If we consider the protocol samples to be equivalent, we can
compare the average performance of the GI protocol (71\%) with the
average performance in the IG protocol (74\%). These two scores are 
statistically the same. Most of the following discussion
focuses on the IG protocol in which it is possible to compare the first individual performance before the group intervention with the second individual performance two weeks after the group intervention.

There are some interesting trends in the time that students took to complete the test during
the two successive tests in the IG and GI protocols.  In the GI
protocol, during the group work only, and in the IG protocol, both
during the individual and group work, students roughly took the whole time alotted to them. In contrast,
students working individually after the group work in the GI protocol
took roughly one third of the time spent on group work. In the IG
protocol, despite having worked on the problems individually, it is
likely that  students were willing to spend  time discussing the
same test because they found peer collaboration useful. Discussions
with individual students support this
hypothesis. However, in the GI protocol, after having discussed the
test with peers, students were reasonably sure about their answers and
did not consider it necessary to reconsider their choices.

\subsection{Evidence for co-construction}

Although there is no consensus in the research literature on the
definition of ``co-construction,''\cite{collaborate} we
use the term here to denote cases where neither student alone choose the
correct response, but both students as a group choose the correct
response. Co-construction can occur for several reasons. For example,
if the group members chose   incorrect responses, they
will have to explain their reasoning to each other. This discussion may
reveal flaws in their initial logic, and complementary information
provided by their peers can help students converge to the correct
solution. Even in cases where both students have the same incorrect
response, co-construction can occur if students are unsure about their
initial response and are willing to discuss their doubts.
Important clues provided by peers during the discussion can trigger
the recall of relevant concepts and can help the group co-construct. One
attractive feature of peer collaboration is that because both peers
have recently gone through similar difficulties in assimilating and
accommodating the new material, they often can  relate to each other's
difficulties more easily than the instructor. The instructors'
extensive experience often can make a concept so obvious and automatic
that they may not comprehend why students misinterpret various concepts
or find them confusing.

Another possible reason for co-construction of knowledge during peer
collaboration is related to reduction in the cognitive
load.\cite{sweller} Cognitive load during problem solving is
the amount of mental resources required to solve the
problem. Cognitive research suggests that the human ``working memory" can keep only seven to nine knowledge pieces (chunks) at a given time during problem solving.\cite{simon} Because students' knowledge structure is more fragmented,
their ``knowledge chunks'' are smaller than that of experts.\cite{simon} For example, displacement, velocity and acceleration may constitute three separate ``chunks" for a beginning physics student, but they form a single knowledge chunk for an expert in mechanics. The limited processing capacity
of the brain makes the
cognitive load high during problem solving tasks, leaving few cognitive
resources available for learning, extending, and organizing
knowledge.\cite{simon} The abstract nature of the laws of physics and
the chain of reasoning required to draw meaningful inferences makes
these cognitive issues critical. According to the
theory of distributed cognition, an individual's cognitive load can be
reduced by taking advantage of the environment. Peer collaboration can
reduce the cognitive load on each individual because the load is shared
by collaborators. Collaborators can take advantage of each other's
strength and the total number of available ``chunks" in working memory is
larger.\cite{simon} 

Table~\ref{2a} displays the percentage of overall cases where neither,
one, or both group members chose the correct response individually, and
how their choices changed during the group work. Co-construction was
observed in 29\% of the eligible cases. Table~\ref{2b} shows the
fraction of responses on each question of the CSEM\cite{csem} that went
from both incorrect individually to correct group response (001:
individual incorrect-individual incorrect-group correct), both
incorrect individually to incorrect group response (000), both correct
individually to correct group response (111), one correct and one
incorrect individually to incorrect group response (100), and one
correct and one incorrect individually to correct group response (101).
The fraction of responses that went from both correct individually to
incorrect group response (110) is negligible. In Table~\ref{2b}, an
incorrect response is labeled 0 regardless of which incorrect choice it
was. Table~\ref{2b} shows that in some questions, the probability of
co-construction was higher than in others.\cite{csem} To determine
whether students were likely to have chosen an incorrect response as
individuals but the correct response as a group due to random guesses,
we analyze the first row of Table~\ref{2a} in detail. In Table~\ref{3},
we subdivide this row based on whether both partners had the same or
different incorrect responses and if the group response was one of the
original incorrect responses or a third incorrect response.
Table~\ref{3} shows that in 25\% of the cases where both group members
had the same incorrect response, and in 31\% of the cases where
both had a different incorrect response, the group response was
correct. In comparison, the relatively small frequency of a
``different'' incorrect group response that was not originally selected
by either member suggests that students were not merely guessing (see
Table~\ref{3}). For example, in the first row of Table~\ref{3} there are
three other well-designed distracters apart from the one originally
chosen by students. If students were randomly guessing, they would have
chosen on average the three other distracters ($0^\prime$) with three
times the probability than they chose the correct response. On the
contrary, only in $8\%$ of the cases in which both students had the
same incorrect response, did they chose
$0^\prime$, compared to choosing the correct response  in 25\% of
the cases.

Although we did not conduct formal interviews with students after they
worked in groups, we briefly discussed aspects of the group work they
found helpful with several students. Most students said that they
obtained useful insights about various electricity and magnetism
concepts by discussing them with peers. Students frequently noted
(often with examples) that they had difficulty interpreting the
problems alone, but interpretation became easier with a peer. They also
said that talking to peers forced them to think more about the
concepts, find fault with their initial reasoning, and remind them of
concepts they had difficulty recalling on their own. Qualitative
observations show that students were more likely to draw field lines,
write equations or sketch drawings in  group work than in their
individual work.

\subsection{Possible negative impacts of unfacilitated peer
collaboration}

Negative impacts of unfacilitated peer
collaboration in introductory physics are possible. For example, a
student with a dominant personality might convince others that his/her
logic is correct, even if it is not. Table~\ref{2a} (second row) shows
that  in the cases in which one individual chose
the correct response  and the other
chose an incorrect response in the IG protocol, 78\% of the group responses were correct. The fact
that 22\% of such cases resulted in an incorrect group response is not
very troublesome because it was an unguided peer discussion. It can
happen if students who individually chose the correct response are not
very confident and cannot defend or justify their response. On most
test questions, when one group member individually had the correct
response and the other had an incorrect response, the group was very
likely to choose the correct response. Individual discussion with
some students also suggests that students who chose an incorrect
response individually were correspondingly less sure and were
more willing to agree with their peer's arguments. The last row of
Table~\ref{2a} shows that when both students individually answered a
question correctly, the group discussion did not result in an incorrect
response to within two significant digits. As will be discussed in
Sec.~\ref{sec4}, group work always resulted in a significantly better
average individual gain in comparison to  no group work.

\section{Individual gain and retention with group
intervention}\label{sec4}

The {individual} performances in the GI protocol was superior (70\%) compared to the IG protocol (56\%). We could
hypothesize that students could immediately recall the group responses
for all 32 test questions in the GI protocol and their superior 
performance does not reflect the effectiveness of group work with
regard to the retention of the concepts discussed. Similarly, in the IG
protocol, the superior group performance compared to the individual
performance is due to a large number of cases where the group member
with the correct response was able to convince the one with the
incorrect response. It does not necessarily imply that students will
retain what they learned in the group work. To investigate the impact
of group interaction on retention,  students took the CSEM test
again two weeks after the IG and GI
protocols. Although it would have been better to use a
different, equally reliable test for assessing similar concepts, none
was available.

The average normalized gain $g$ can be defined as
$g =\langle (s_f-s_i)/(100-s_i) \rangle$, where $s_i$ and $s_f$ are
the first individual and second individual test scores in
percent.\cite{hake} All of the gains in the various tables
were calculated before rounding the first and second average
individual scores to two significant digits. For the students in IG protocol, the
average score for the second individual testing was 74\% (the same as
the average group score earlier), a gain of
0.41 compared to the average initial individual score of 56\%. A
detailed comparison of the group and second individual test scores
shows that about 80\% of the overall individual responses chosen by
the members of a particular group were the same as the group
responses. There are two competing effects: the fact that students
forgot some group responses  and the fact that they had two weeks to
study the questions before the second individual test. Thus, a large
fraction of the group responses was retained even after two weeks. We
note that students did not know that they would be taking the CSEM
test again. The average second individual score for the GI students
protocol was 70\%. Although this score is lower than the average
second individual score for the IG students  (74\%), the difference is
not statistically significant.

\subsection{Effective pairing}

To learn about the effect of pairing students with different initial
individual scores, we divided the 148 students in the IG protocol 
into three categories: high (A), middle (B), and low 
(C), based on their initial individual scores on the CSEM. All
students with an initial score $\geq 70\%$
were placed in category A, those with scores less than 70\% but
greater than 50\% were placed in category B and those with scores
$\leq 50\%$ were placed in category C. In
Tables~\ref{4a} and
\ref{4b} we show the  initial individual average score and the second
individual score in each category for all nine possible pairs. The top
rows of  Tables~\ref{4a} and \ref{4b} refer to the performance of A
students for different kinds of pairings: (A\,A), (A\,B) and (A\,C).
Similarly, the middle and bottom rows refer to the performance of B and
C students for different kinds of pairings respectively. Table~\ref{4a}
shows that the  initial average scores of the students in the A,
B, and C categories regardless with whom they were paired 
were $\approx 77\%$, 58\%, and 40\% respectively. Table~\ref{4b} shows
that the  second individual average scores of the students in the 
A, B, and C categories regardless with whom they were paired
were $\approx 86\%$, 75\%, and 64\% respectively. In comparison, for
all 148 students together, the  initial individual average score was
56\%, the average individual gain was 0.41, and the  second
individual average score was 74\%. The difference in the normalized
gain in different categories in the matrix is not statistically
significant.

\section{Comparison with students without group intervention}
To gauge the effectiveness of group work on individual performance, we
employed a control group in which 178 students from two different
calculus-based introductory physics courses took the CSEM test once
individually (average score 57\%) and then again two weeks later
without any group intervention (average score 63\%). The normalized
gain for the control group is 0.14, which is much less than the gain
of 0.41 for students in the IG protocol who worked in pairs before
the individual testing two weeks later. Table~\ref{5} shows the
average first (I) individual score, the second (II) individual score,
and the normalized gain
$g$ for the control students divided into the same A,  B, and C
categories according to their first individual score and for all
students. Table~\ref{5} shows that the performance of the control
group in none of the categories was much improved two weeks later. A
comparison of Tables~\ref{4b} and
\ref{5} shows that in each category students in the IG protocol
obtained roughly 10 points higher on the second individual testing than
those in the control group.

\section{Summary}

We have investigated the effectiveness of working in pairs without
facilitation from the instructor on the CSEM test in a calculus-based
introductory physics course. Students who worked with peers showed
significantly higher normalized gain on subsequent individual testing
than a control group that took the test individually twice. In our
limited sample, we did not find any statistical difference between the
performance of students in the IG and GI protocols.

The peer
collaboration also shows evidence for co-construction. Students who
individually chose an incorrect response were able to find the
correct response working as a group with a frequency that is roughly
ten times higher than that predicted by random guessing.
Co-construction also happened in cases where the individual incorrect
responses were the same. Discussions with individual students show
that discussing their doubts with each other helped them.
Peer collaboration provided students an opportunity to articulate
their own thoughts and make sense of their peer's thought processes.
This process made students critical of their own thinking. Discussions
with individual students indicated that they 
value peer interaction. Also, students who had worked on the test
individually, when asked to work with peers immediately following the
individual work, used the entire time allotted to them (in contrast to
the group that first worked with peers and then individually).
We found no
significant differences between individual gains regardless of
whether a student with a high individual score was paired with
another student who had a high or a low individual score.  

Because unfacilitated peer collaboration requires a minimal effort on
the part of instructors, students should be given ample opportunity and
incentive to collaborate with peers both inside and outside of the
classroom. There are several factors that seem to be important for
optimizing the benefits of peer collaboration. A time constraint, even
for the collaborative work done outside of the classroom (for example,
a time frame within which the work should be submitted) may be helpful
for keeping students focussed. A reward system (for example, a small
amount of homework, quiz, or bonus points), and an incentive for
individual accountability (for example, future examination in which
students will work alone on similar concepts) is helpful for getting
the most out of collaborative work.

\begin{acknowledgments}
We are very grateful to M.\ T.\ H.\ Chi, R.\
Glaser, R.\ Johnsen, and J.\ Levy for useful discussions. This work is
supported in part by the National Science Foundation (award PHY-0244708).
\end{acknowledgments}

\newpage

\section*{Tables}

\begin{table}[h]
\centering
\begin{tabular}[t]{|c|c|c|c|}
\hline
 Protocol &I&G \\[0.5 ex]
\hline \hline GI& 70\% & 71\%\\[0.5 ex]
\hline IG& 56\% & 74\% \\[0.5 ex]
\hline
\end{tabular}
\caption{\label{table1}The average individual (I) and group (G)
scores for the GI and IG protocols. There were 148 students in the IG
protocol and 84 students in the GI protocol.}
\end{table}

\begin{table}[h]
\centering
\begin{tabular}[t]{|c|c|c|c|}
\hline
\multicolumn{2}{|c|}{Individual Response } &\multicolumn{2}{|c|}{Group
Response}{}\\[0.5 ex]\cline{3-4}
\multicolumn{2}{|c|}{} &Incorrect&Correct\\[0.5 ex]
\hline \hline neither correct& 24\% & 71\% & 29\% \\[0.5 ex]
\hline one correct& 40\% & 22\% & 78\% s\\[0.5 ex]
\hline both correct& 36\% & 0\% & 100\% \\[0.5 ex]
\hline
\end{tabular}
\caption{\label{2a}Distribution of the average group response
for various combinations of individual responses of group members in
the IG protocol. The second column displays the percentage of the
overall cases where neither, one, or both group members had the
correct response individually.}
\end{table}

\newpage
\baselineskip=20pt
\begin{longtable}[t]{|c|c|c|c|c|c|}
%\centering
%\begin{tabular}[t]{|c|c|c|c|c|c|}
\hline
Question $\#$&001&000&111&100&101\\
\hline
 1 & 0.00 & 0.01 & 0.92 & 0.00 & 0.07\\
 2 & 0.01 & 0.08 & 0.73 & 0.05 & 0.12\\
 3 & 0.00 & 0.04 & 0.84 & 0.01 & 0.11\\
 4 & 0.04 & 0.08 & 0.43 & 0.12 & 0.32\\
 5 & 0.03 & 0.09 & 0.39 & 0.07 & 0.42\\
 6 & 0.01 & 0.03 & 0.73 & 0.04 & 0.19\\
 7 & 0.08 & 0.14 & 0.26 & 0.09 & 0.43\\
 8 & 0.00 & 0.01 & 0.69 & 0.04 & 0.26\\
 9 & 0.04 & 0.00 & 0.42 & 0.12 & 0.42\\
 10 & 0.07 & 0.11 & 0.24 & 0.07 & 0.51\\
 11 & 0.11 & 0.27 & 0.22 & 0.08 & 0.32\\
 12 & 0.03 & 0.03 & 0.66 & 0.04 & 0.23\\
 13 & 0.03 & 0.16 & 0.34 & 0.08 & 0.39\\
 14 & 0.05 & 0.38 & 0.19 & 0.14 & 0.24\\
 15 & 0.11 & 0.20 & 0.26 & 0.08 & 0.35\\
 16 & 0.09 & 0.11 & 0.28 & 0.12 & 0.39\\
 17 & 0.03 & 0.16 & 0.36 & 0.14 & 0.31\\
 18 & 0.12 & 0.23 & 0.22 & 0.20 & 0.23\\
 19 & 0.18 & 0.15 & 0.27 & 0.08 & 0.32\\
 20 & 0.18 & 0.43 & 0.03 & 0.18 & 0.19\\
 21 & 0.04 & 0.38 & 0.12 & 0.07 & 0.39\\
 22 & 0.09 & 0.23 & 0.16 & 0.23 & 0.28\\
 23 & 0.09 & 0.07 & 0.62 & 0.01 & 0.20\\
 24 & 0.08 & 0.27 & 0.24 & 0.07 & 0.34\\
 25 & 0.03 & 0.08 & 0.42 & 0.08 & 0.39\\
 26 & 0.05 & 0.03 & 0.55 & 0.00 & 0.36\\
 27 & 0.03 & 0.28 & 0.15 & 0.15 & 0.39\\
 28 & 0.08 & 0.12 & 0.27 & 0.08 & 0.45\\
 29 & 0.14 & 0.18 & 0.20 & 0.11 & 0.38\\
 30 & 0.08 & 0.20 & 0.34 & 0.15 & 0.23\\
 31 & 0.14 & 0.45 & 0.08 & 0.03 & 0.31\\
 32 & 0.15 & 0.43 & 0.05 & 0.14 & 0.23\\
\hline
%\end{tabular}
\caption{\label{2b}In the IG protocol, the fraction of
responses on each question that went from both incorrect individually
to correct group response (001), both incorrect individually to
incorrect group response (000), both correct individually to correct
group response (111), one correct and one incorrect individually to
incorrect group response (100) and one correct and one incorrect
individually to correct group response (101). Instances where both
correct individually went to incorrect group response (110) were
negligible.}
\end{longtable}

\begin{table}[h]
\centering
\begin{tabular}[t]{|c|c|c|c|}
\hline both individual &\multicolumn{3}{|c|}{Group
Response}{}\\[0.5 ex]\cline{2-4} responses were incorrect&1&0
&$0^{\prime}$\\[0.5 ex]
\hline \hline same incorrect (41\%) & 25\% &67\% & 8\%\\[0.5 ex]
\hline different incorrect (59\%) & 31\% & 55\% &14\%\\[0.5 ex]
\hline
\end{tabular}
\caption{\label{3}Distribution of the average group response
for cases where both members had the same or different incorrect 
individual response in the IGI protocol.
$1$, $0$ and $0^\prime$ refer to ``correct,'' ``one of the original
incorrect'' and ``an incorrect choice not originally selected by either
student'' group response respectively.}
\end{table}

\begin{table}
\centering
\mbox{
\begin{minipage}{0.3 \linewidth}
\begin{tabular}[t]{|c|c|c|c|}
\hline (a)&\multicolumn{3}{|c|}{pairing}\\[0.5 ex]\cline{2-4} &A &B &C
\\[0.5 ex]
\hline \hline A&78&77&75\\[0.5 ex]
\hline B&58&59&58\\[0.5 ex]
\hline C&43&39&39\\[0.5 ex]
\hline
\end{tabular}
\end{minipage} }
\caption{\label{4a}The average initial individual score in
percent. The top row
refers to the performance of A students for different kinds of
pairings: (A\,A), (A\,B) and (A\,C). Similarly, the middle and bottom
rows refer to the performance of B and C students, respectively. There
were a total of 148 students out of which 12 chose AA pairing, 12
chose
BB pairing, 24 chose CC pairing, 36 chose AB pairing, 20
chose AC pairing and 44 chose AC pairing.}
\end{table}

\begin{table}[h]
\centering
\begin{tabular}[t]{|c|c|c|c|}
\hline &\multicolumn{3}{|c|}{pairing}\\[0.5 ex]\cline{2-4} &A &B &C
\\[0.5 ex]
\hline \hline A&88&86&85\\[0.5 ex]
\hline B&79&71&75\\[0.5 ex]
\hline C&64&65&62\\[0.5 ex]
\hline
\end{tabular}
\caption{\label{4b}The average second individual test score in
percent for the nine types of pairs in the IG protocol} 
\end{table}

\begin{table}[h]
\centering
\begin{tabular}[t]{|c|c|c|c|}
\hline {Student Type} &I&II&g\\[0.5 ex]
\hline \hline A&73&77&0.14\\[0.5 ex]
\hline B&58 &65&0.17\\[0.5 ex]
\hline C&42&49&0.12\\[0.5 ex]
\hline All&57&63&0.14\\[0.5 ex]
\hline
\end{tabular}
\caption{\label{5}The percent average first (I) individual
score, the second (II) individual score two weeks later, and the
normalized gain $g$ for the 178 control students (no group intervention)
divided in high ($A$), middle ($B$) and low ($C$) categories according
to their first individual score and for all students. There were 56
students in category $A$, 54 in category $B$ and $68$ in category $C$.
The gains were calculated before rounding the initial and second
average individual scores to two significant digits.}
\end{table}

\end{document}